\documentclass[]{raa}
\usepackage{graphicx,times}
\usepackage{natbib}

\date {}

\begin{document}
\title{Pulse profile stability of the Crab pulsar}
   \setcounter{page}{1}

   \author{Chetana Jain
      \inst{1}
   \and Biswajit Paul
      \inst{2}
   }

   \institute{Hans Raj College, University of Delhi, Delhi-110007, India; {\it chetanajain11@gmail.com}\\
	\and
	Raman Research Institute, Sadashivnagar, C. V. Raman Avenue,  
             Bangalore 560080, India.
}

\abstract{We present an X-ray timing analysis of the Crab pulsar, PSR B0531+21, using the archival $RXTE$ data. We have investigated the stability of the Crab pulse profile, in soft (2$-$20 keV) and hard (30$-$100 keV) X-ray energies, over the last decade of $RXTE$ operation. The analysis includes measurement of the separation between the two pulse peaks; and intensity and the widths of the two peaks. We did not find any significant time dependency in the pulse shape. The two peaks are stable in phase, intensity and widths, for the last ten years. The first pulse is relatively stronger at soft X-rays. The first pulse peak is narrower than the second peak, in both, soft- and hard X-ray energies. Both the peaks show a slow rise and a steeper fall. The ratio of the pulsed photons in the two peaks is also constant in time. 
\keywords{pulsars: individual: Crab (PSR B0531+21) -- stars: neutron -- X-rays
}
}
   \authorrunning{C. Jain \& B. Paul}            
   \titlerunning{Pulse profile stability of the Crab Pulsar }  
   \maketitle

\section{Introduction}
\label{sect:intro}
The Crab pulsar (PSR B0532+21) was discovered in 1968 \citep{Staelin68}, and is the remnant of a supernova explosion (SN 1054). The neutron star rotates about 30 times per second. The pulsed emission is visible across the entire electromagnetic spectrum (radio: \citet{Hankins07}, optical: \citet{Oosterbroek08}, X-rays: \citet{Mineo06}, $\gamma$-rays: \citet{Kuiper01}, and very high energy: \citet{Aliu08}).

In all the energy bands, the Crab pulse profile exhibits a dual peak structure and the pulse morphology is known to vary as a function of photon energy. The two peaks are separated by a phase of $\sim$ 0.4 over all the wavelengths \citep{Pravdo97, Kuiper03, Rots04}. But arrival times of the two peaks are not aligned in phase at different wavelengths. X-ray and $\gamma$-ray pulse leads the radio pulse \citep{Kuiper03, Rots04, Abdo10}. The phase delay measurements between optical and radio observations show a fuzzy behaviour, with reports of optical pulse leading \citep{Oosterbroek06}, trailing behind \citep{Golden00} and being coincident \citep{Romani01}, with the radio pulse. 

The variation of Crab pulse profile as a function of photon energy has been studied umpteen times with significant detail. The relative intensity of the two pulse peaks move in see saw manner with increasing energy. The first peak is more pronounced at radio and optical wavelengths \citep{Oosterbroek08}. The second peak starts dominating as energy increases; and eventually towers above the former at soft $\gamma$-rays \citep{Mineo08}. With a further increase in energy, the first peak again becomes dominant \citep{Kuiper01}. 

The time variability of the Crab pulse profile is relatively less studied \citep{Jones80, Carraminana94}. Nevertheless, it is equally important. It can assist our understanding of the emission nature of this object. Irrespective of the physical internal process in the neutron star, phenomena that can possibly change the high energy pulse profile of a rotation powered pulsar are:
(i) some reconfiguration of the neutron star crust and magnetic field structure, which can be associated with starquaks and glitches \citep{Franco00} and
(ii) changes in the neutron star spin axis with respect to the line of sight, which can happen due to precession \citep{Melatos00}.
Some variations of the gamma-ray pulse profile of Crab was interpreted to be due to a 13 year precession \citep{Nolan93, Ulmer94}. Pulsar glitches are expected to produce thermal X-rays \citep{Hui04} and certain form of energy release can also produce pulsed X-ray emission \citep{Tang01, Cheng98}, which can change the pulse profiles for some period after the glitches.

In this paper, we present the first ever long term pulse shape analysis of the Crab pulsar, using data spread over 10 years (2001-2010). Observations made with the instruments on board the $RXTE$ satellite, were used to analyze the evolution of the pulse profile in $-$ soft (2$-$20 keV) and hard (30$-$100 keV) X-ray energy band. In section 2, we describe the data reduction from the $RXTE$-PCA and HEXTE instruments. This section also explains the model used to quantify the observations. The results obtained from the detailed timing analysis are discussed in Section 3. We summarize and discuss the implications of the observations in Section 4. 

\section{Observations and analysis}
\label{sect:Obs}

Data for the present analysis were obtained from two different detectors on board the $RXTE$: (i) the Proportional Counter Array (PCA), and (ii) the High-Energy X-ray Timing Experiment (HEXTE). The field of view of these non-imaging detectors contained both the Crab pulsar and the Crab nebula. The data used in the present analysis are in the 2-20 keV band from the PCA detectors and in the 30-100 keV band from HEXTE instrument. Crab nebula suffered significant scattering during 1995-1998 \citep{Wong01, Rots04}. Therefore, we did not include data before this period, so as to avoid misleading results due to broadening of the pulses. The entire analysis was done using the \textsc{ftool} from the astronomy software, \textsc{heasoft}-ver 5.2. The following subsections give the details of PCA and HEXTE data reduction.

\subsection{Reduction of $RXTE$-PCA data}

The $RXTE$-PCA consists of an array of five collimated xenon/methane multi anode proportional counter units (PCU) with a total photon collection area of 6500 cm$^{2}$ \citep{Jahoda96, Jahoda06}. Data is available in different event modes and they are binned into different energy channels. For the present work, we have used all the publicly available data in event mode, E$_{-}$250$\mu$s$_{-}$128M$_{-}$0$_{-}$1s. This data mode has a time resolution of 250 $\mu$s and energy information spread over 128 channels. Selection of this mode, allowed us a uniformity over the energy bins over a long timebase from 2001--2010. 

From each observation, energy resolved pulse folded light curves were generated using the \textsc{ftool}-\textsc{fasebin}. This tool corrects the photon arrival times and converts them to the solar system barycenter. The arrival time of each event is then converted into the phase of rotational period of the pulsar. The output of \textsc{fasebin} is a two dimensional histogram of photon count rate against the pulse phase and energy. Absolute phase was determined by using the time solution available from the Jodrell Bank Crab Pulsar Monthly Ephemeris\footnote[1]{http://www.jb.man.ac.uk/} \citep{Lyne93} and the corresponding pulsar coordinates.
%The database is available at $``$http://www.jb.man.ac.uk/".
Thereafter, another \textsc{ftool} - \textsc{fbssum} was used, which integrates the pulse profiles over different energy ranges.

A typical 2$-$20 keV $RXTE$-PCA pulse profile, binned into 128 phasebins, is shown in Figure 1 (left Panel). The figure shows the average profile of all the observations made between January to June, 2009. The average profile was generated using the \textsc{ftool}-\textsc{fbadd}. The folded light curve has an asymmetric, dual-peak shape. The first pulse is stronger than the second pulse. A significant inter peak emission (bridge emission) is also seen. Hereafter, first peak will be referred to as $``$P1" and the second peak as $``$P2".

\begin{figure}
\centering
\includegraphics[height=2.8in, width=2.7in, angle=-90]{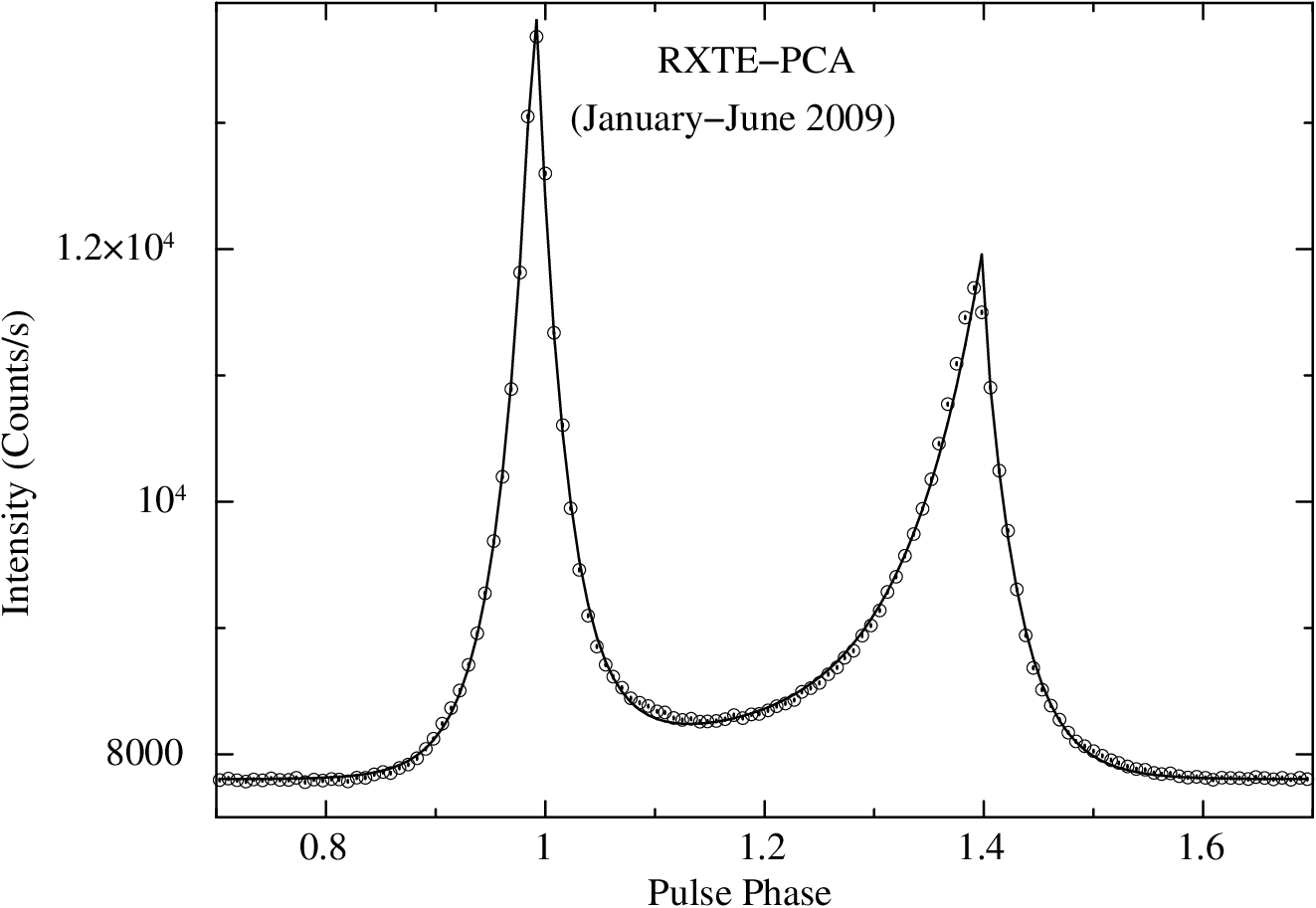}
\includegraphics[height=2.8in, width=2.7in, angle=-90]{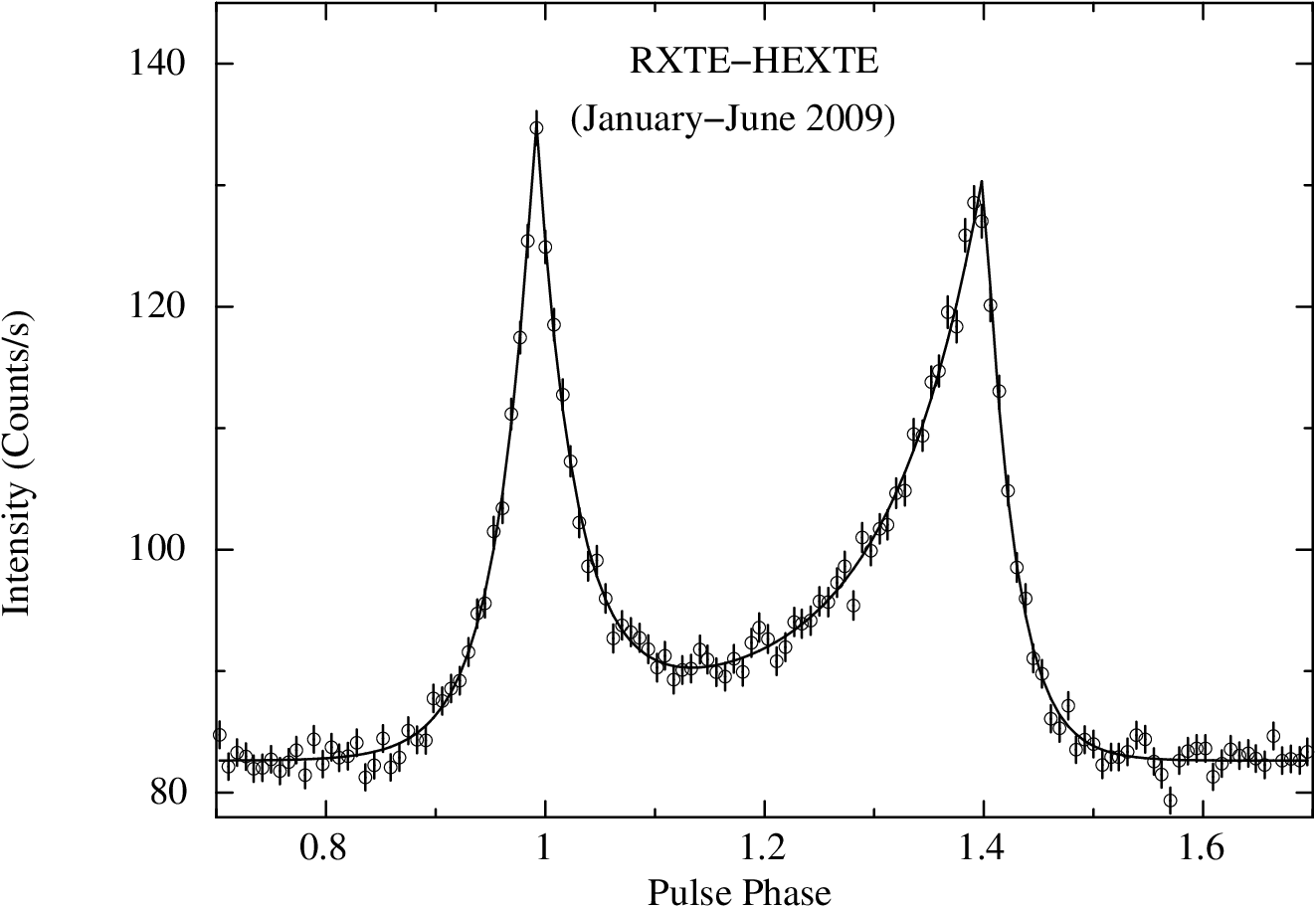}
\caption{A sample of 2$-$20 keV and 30$-$100 keV, pulse profiles of Crab pulsar, from $RXTE$-PCA and HEXTE observations. The solid line shows the best fit model, with a reduced $\chi^{2}$ of 1.3 and 0.9, for 118 d.o.f., respectively. The observation details are mentioned in each panel.}
\end{figure}

\subsection{Reduction of $RXTE$-HEXTE data}

The $RXTE$-HEXTE consists of 2 clusters each containing four NaI/CsI phoswich scintillation detectors, with a total photon collection area of 1600 cm$^{2}$ \citep{Rothschild98}. Each of these clusters rock along mutually orthogonal directions to provide background measurements. When cluster-0 is on target, cluster-1 is off target and vice-versa, i.e. the HEXTE rocks between on- and off- source positions. For the present work, data in science event mode was analyzed from all the observations made between 2001-2010. From the raw science event files, the source file were generated, using the \textsc{ftool}-\textsc{hxtback}. Energy resolved pulse profiles were obtained using the \textsc{ftools}- \textsc{fasebin} and \textsc{fbssum}. The \textsc{fasebin} extracts only the on-source data from the raw file. The HEXTE data was also folded with the same radio ephemeris, as in case of PCA data. 

Figure 1 (right panel) shows a sample of average 30$-$100 keV pulse profile of Crab pulsar, from data obtained between January-June, 2009. As opposed to the soft X-ray pulse profile, the data above 30 keV has low statistics. We therefore, separately, generated HEXTE pulse profiles by averaging the data on half yearly basis.

\subsection{Model fitting}
To quantify the analysis, we fitted a 10-parameter model to the pulse profiles. The model consisted of position and peak intensity of each pulse peak; 2 exponentials (1 rise and 1 decay) for each pulse peak; and 2 constants (one for the nebular and background emission; and the other for the inter-peak emission). The model was fit over the phase range 0.7 -1.7. An example of the best fit model is shown with a solid line, in Figure 1, for both - PCA and HEXTE profiles. The reduced $\chi^{2}$ of the best fit ranged between 0.6 and 3.5, for 118 degrees of freedom.

\section{Results}

All the pulse profiles of $RXTE$-PCA and HEXTE, were fitted with the model as described above. The variability/stability of the important parameters of the fit, are discussed below. Several glitches have been detected from the Crab pulsar in the period of these X-ray data. The times at which the glitches occured (taken from \citet{Espinoza11}) are marked in the top panel of each subsequent figure. All except one glitch had similar magnitudes ($\Delta\nu/\nu$ of $\sim$10 $\times$ 10$^{-9}$), while one of them (which occured around MJD 53067) had a magnitude of $\sim$200 $\times$ 10$^{-9}$.

\subsection{Separation of the two pulse peaks}

As shown in the sample pulse profiles in Figure 1, two clear peaks are seen at phases $\sim$0.99 and $\sim$1.39. Figure 2 shows the separation between the two peaks, from the $RXTE$-PCA and HEXTE observations. The X-axis in this figure, gives the MJD of respective observation, and the Y axis gives the phase separation between the two peaks.

\begin{figure}
\centering
\includegraphics[height=3.5in, width=2.6in, angle=-90]{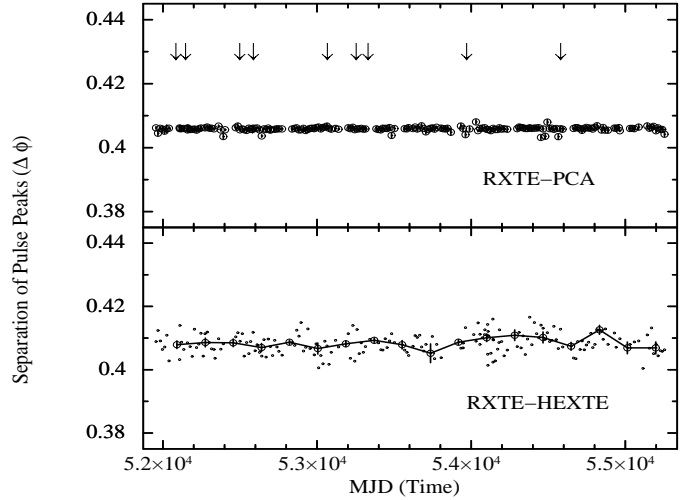}
\caption{The separation between the two pulse peaks of the Crab pulse profile, from $RXTE$-PCA and HEXTE observations. The solid line in the bottom panel shows the half yearly trend in HEXTE. Arrows in the top panel indicate the time of occurence of glitches.}
\end{figure}

A constant fit to the $RXTE$-PCA curve gave a phase separation of 0.4058 (5), with a reduced $\chi^{2}$ of 0.94 for 189 d.o.f. A constant fit to the HEXTE results, gave a phase separation of 0.4079 (6), between the two pulse peaks. The reduced $\chi^{2}$ of fit was 0.36 for 188 d.o.f. The solid line in the results from HEXTE detector, show the half yearly trend. This indicate that the pulse peak maximas are stable in time. The separation of the maxima of the two peaks is constant over the last ten years. Further, the phase separation between the two peaks do not show a significant variation with energy, consistent with earlier observations \citep{Wills82, Clear87, Nolan93}. We did not find any considerable effect of occurence of glitches on the separation between the pulse peak maximas.

\subsection{Pulse widths}

The rise times and the decay times of P1 and P2 (in phase units), for the PCA and HEXTE pulse profiles are shown in Figure 3, along with occurence of glitches in the top panel. The X-axis gives the MJD of each observation. In case of HEXTE, the half yearly average trend is shown with a solid line. A constant was fit and results are given in Table 1, with 90\% confidence limit, and $\chi^{2}$ of fitting. In both the energy bands, the two pulses are asymmetric. Both the peaks, P1 and P2, show a slower rise and a steep fall. The first pulse peak is narrower than the second peak. It rises in almost half the time taken by the second peak. The first peak also decays faster than the second peak. We did not find any detectable effect of occurence of glitches on the pulse widths.

\begin{figure}
\centering
\includegraphics[height=2.8in, width=3.0in, angle=-90]{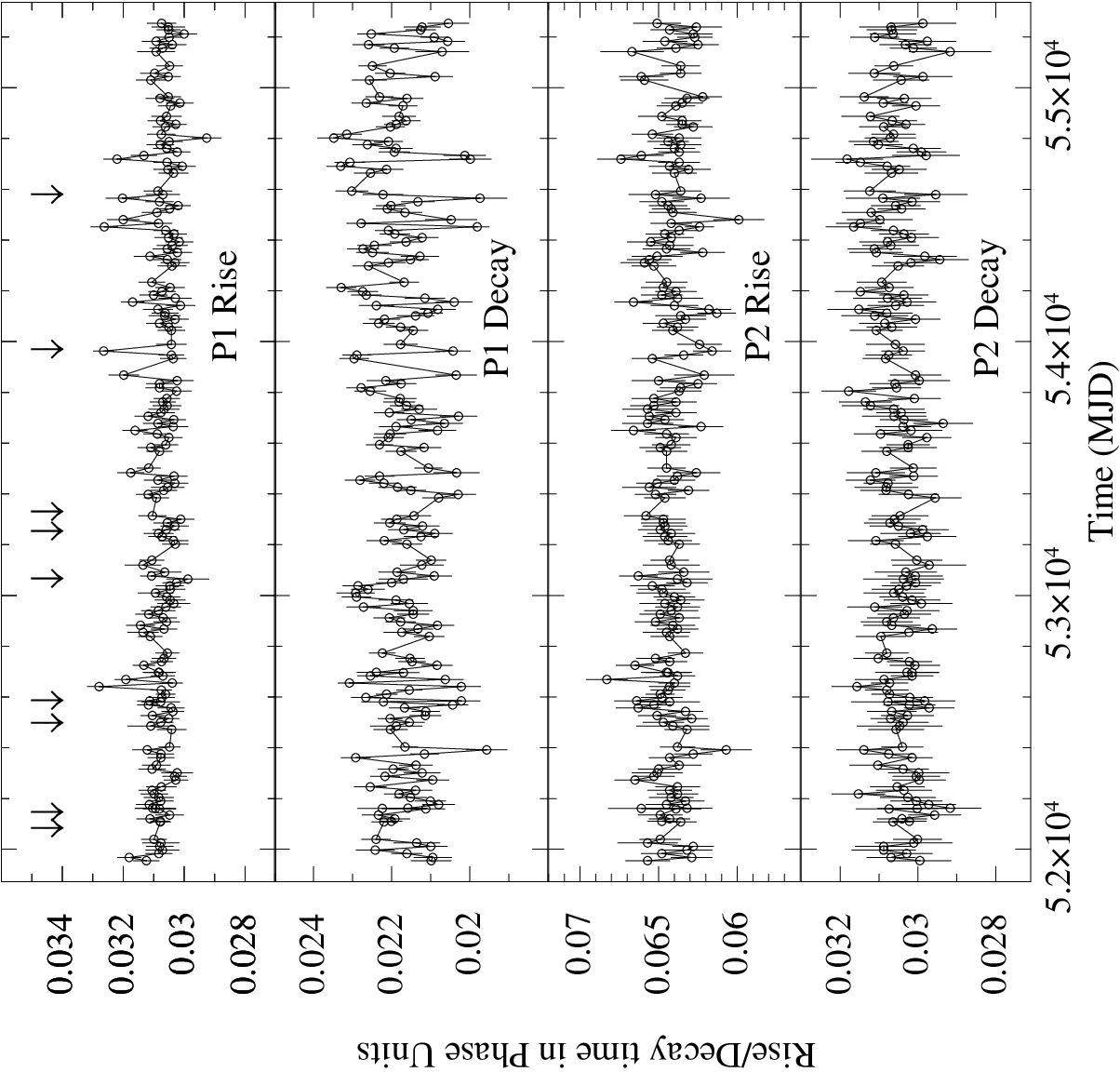}
\includegraphics[height=2.8in, width=3.0in, angle=-90]{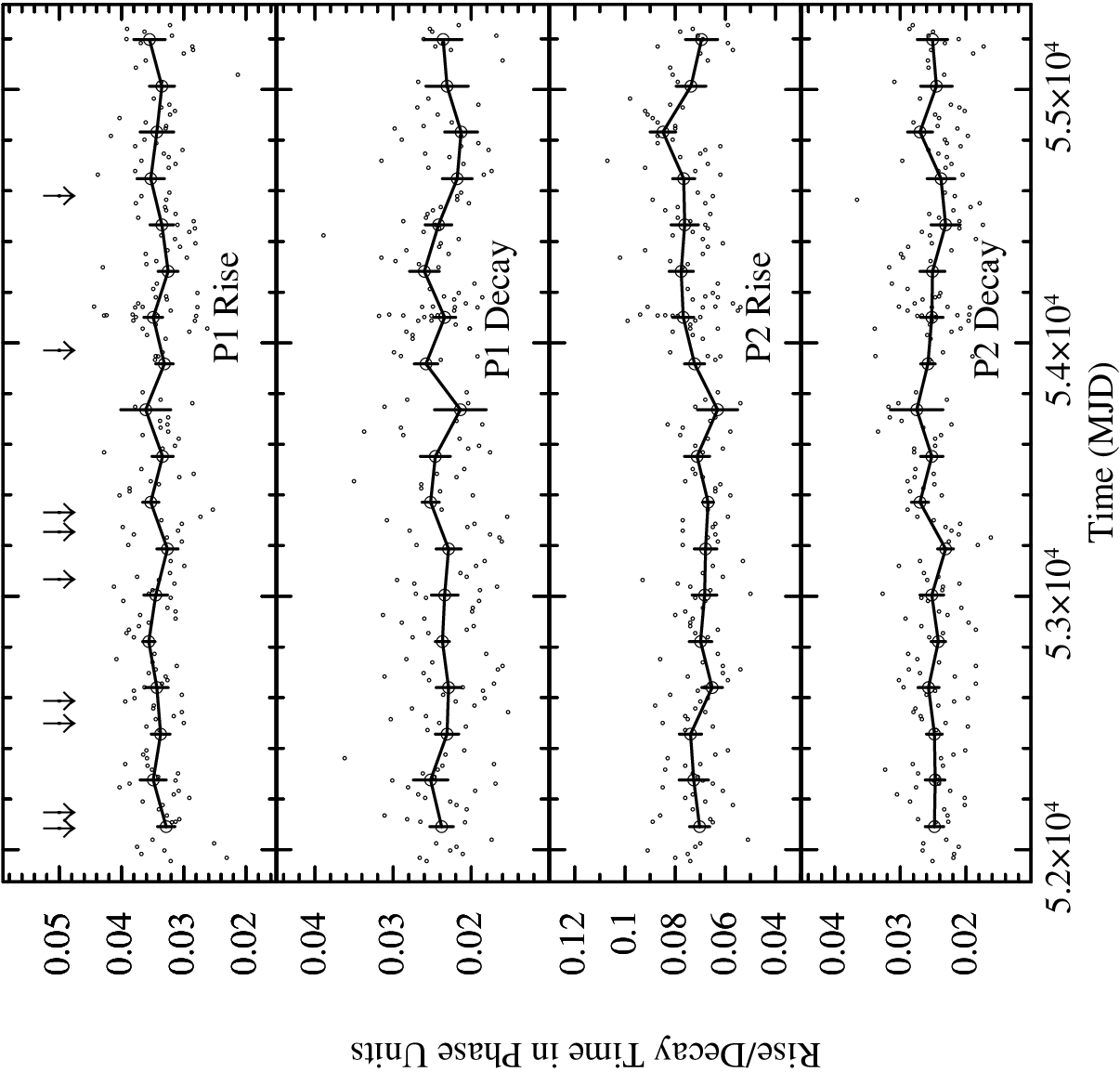}
\caption{The rise and decay times of the two peaks of the Crab pulse profile. The left panels show the results from $RXTE$-PCA observations, while the HEXTE results are shown in the right panels.The half yearly trend of the HEXTE observations is shown with a solid line in the figure. The arrows indicate the time of occurence of glitches.}
\end{figure}

\subsection{Pulsed intensity} 

Figure 4 shows the ratio of the integrated pulsed photons in the two peaks, for $RXTE$-PCA and HEXTE observations. A constant fit to the two ratios, gave a value of 0.633 (1) for PCA and 0.604 (13) for HEXTE energy bands. This result is consistent with results reported using $COMPTEL$ observations at medium $\gamma$-ray energies (1-10 MeV) \citep{Buccheri93, Carraminana94, Kuiper01}; and from $EGRET$ observations above 30 MeV \citep{Fierro95, Tompkins97}. However, more variation is seen in case of HEXTE data, as compared to PCA results, which might be due to poorer statistics in case of HEXTE data. Further, although the second peak is smaller, it contains large number of pulsed photons.

\begin{figure}
\centering
\includegraphics[height=3.6in, width=2.8in, angle=-90]{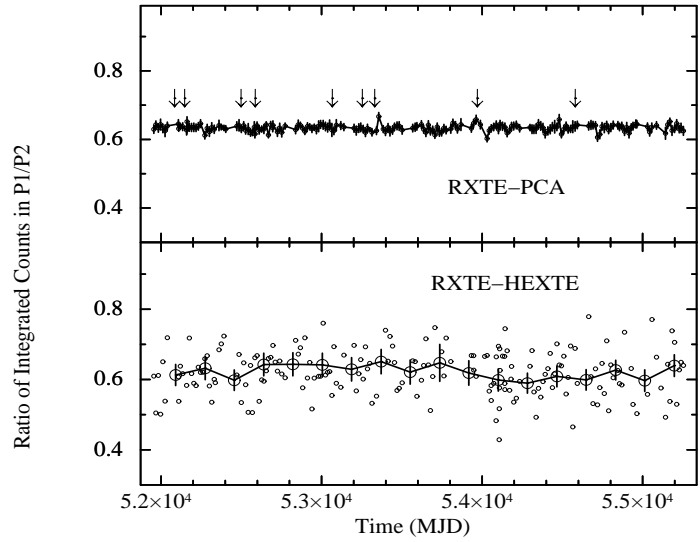}
\caption{The ratio of integrated counts in P1 to P2, for $RXTE$-PCA and HEXTE observations, as a function of time. The arrows indicate the time of occurence of glitches.}
\end{figure}

\subsection{Pulse Intensity Ratio}

A comparison of variation in intensity ratio of the amplitude of the two pulse peaks, in PCA and HEXTE profiles, is shown in Figure 5. Here, the intensity ratio of the two peaks (P1/P2) is plotted as a function of time. A ratio of 1 or above, indicates that the first peak is dominant, whereas a ratio of less than 1 imply a stronger second peak. Clearly, P1 is significantly stronger than P2 in soft X-ray band (PCA data). A constant fit gave an intensity ratio of 1.63, with a reduced $\chi^{2}$ of 0.8 for 189 d.o.f. The intensity ratio of the HEXTE pulses shows a variation, which could be due to poorer statistics. The solid line in the figure, shows the half yearly average trend in the intensity ratio of HEXTE profiles. A constant fit in case of HEXTE results gave an intensity ratio of 1.18, with a reduced $\chi^{2}$ of 0.33 for 188 d.o.f. Clearly, the first peak dominates at higher energy also, albeit the ratio is smaller. 

\begin{figure} 
\centering
\includegraphics[height=3.4in, width=2.5in, angle=-90]{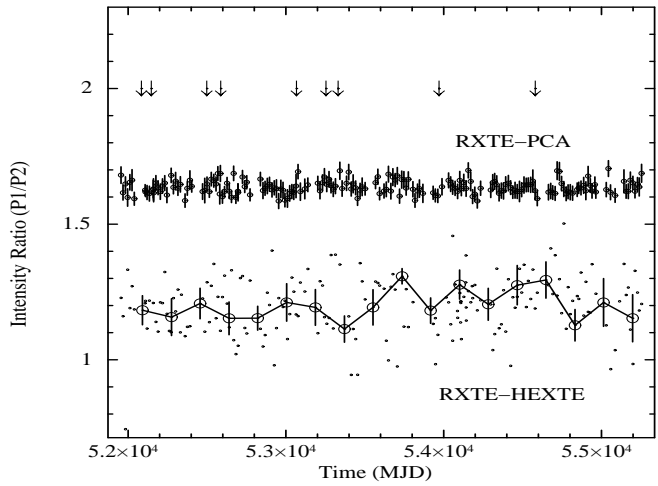}
\caption{The variation of intensity ratios of the two pulse peaks (P1/P2) of Crab pulsar, are shown as a function of time. The curve at the top show the variation for PCA observations; while the curve below corresponds to HEXTE observations. The solid line in the lower curve, shows the half yearly average trend in the HEXTE observations. The arrows indicate the time of occurence of glitches.}
\end{figure}

\begin{table}
\centering
\caption{Results for model fitting of the Crab pulse profile.}
\begin{tabular}{l |c|c|c|c}
\hline
Parameter				&	\multicolumn{2}{c|}{RXTE-PCA}	& \multicolumn{2}{c}{RXTE-HEXTE}\\
\hline
					&Value		&Reduced $\chi^{2}$ 	&Value		&Reduced $\chi^{2}$\\
					&		&(for 189 d.o.f.)	&		&(for 188 d.of.)\\
\hline
Phase separation between two pulse peaks&0.4058 (5)	&	0.94		&0.4079 (6)	&0.36\\
Rise time of P1	(in phase unit)		&0.03077 (5)	&	1.3		&0.0335 (6)	&0.43\\
Decay time of P1 (in phase unit)	&0.02180 (4)	&	3.5		&0.0236 (12)	&2.5\\
Rise time of P2	 (in phase unit)	&0.0642 (1)	&	0.75		&0.0707 (15)	&0.44\\
Decay time of P2 (in phase unit)	&0.03051 (8)	&	0.52		&0.0241 (5)	&0.49\\
Ratio of pulsed photons in two peaks (P1/P2)&	0.633 (1)&	0.63		&0.604 (13)	&0.39\\
Pulse intensity ratio (P1/P2)		&1.6344 (35)	&	0.8		&1.18 (2)       &0.33\\
\hline
\end{tabular}
$\ast$The numbers in bracket, give the error for 90\% confidence range in the respective values.
\end{table}

\section{Discussion}
\label{sect:discussion}

A coherent picture of the pulsed emission from Crab pulsar has been an enigma inspite of numerous multi wavelength studies. The radiation in radio, optical, X-rays and high energy $\gamma$-rays is known to differ considerably. In this paper, we have reported that the X-ray pulse profile of the Crab pulsar is stable in time over the last ten years of $RXTE$ performance. There is no indication of time variability of the pulse shape at soft and hard X-ray energies, during the last decade of observations. No evidence of time variation in the phase separation between the two pulse peaks, the widths of the two peaks and their relative intensities, was noticed. 

The light curves of X-ray pulsars reflect the geometry of the magnetic field and the radiation emission pattern of the rotating object. If the pulses are produced near the neutron star surface, the compactness (mass to radius ratio) of the neutron star will also cause gravitational bending of light and thus effect the pulse profile.
There are several reports on the long-term flux variability of the Crab pulsar. This is conjectured to be due to slow precession of the spin axis of the neutron star, which results in precession of the pulsar magnetosphere at similar periods \citep{Ozel89, Kanbach90, Nolan93, Ulmer94}. Since the X-ray and $\gamma$-ray emission is modulated by the magnetosphere geometry, therefore, changes in the respective pulse profiles should be in sync. Flux variability can also be attributed to sudden glitch-induced changes between the pulsar spin axis and the magnetic moment axis \citep{Link92, Link97, Franco00}. A change in the alignment angle during a glitch, can modulate the duration of the line of sight's traverse through the pulse emission cone. This in turn can produce considerable differences in the pulsed emission from the neutron star \citep{Jones80, Sekimoto95}.

However, if these theories are true, then the pulsed X-ray and the $\gamma$-ray emission are supposed to show similar variations. The $\gamma$-ray pulse profile of the Crab pulsar shows considerable variations over time. The intensity ratio of the two peaks in the pulsar's light curve, is known to undergo a periodic behaviour \citep{Wills82, Ozel91, Nolan93, Ulmer94} at high $\gamma$-ray energies. However, we have found that the intensity ratio of the two peaks is stable in time, both in soft and hard X-rays. This is consistent with earlier works in X-ray band \citep{Wills82}. It has been proposed that the periodic variability in the intensity ratio of the two peaks is masked at lower photon energies, due to onset of plasma-dominated radiation, which mainly affects the second pulse peak \citep{Hasinger84, Hasinger85}. But neither the magnitude of the radiation component is certain, nor are there supportive conclusions from the optical and radio observations.

The emission of thermal radiation from the neutron star is also affected by the turmoil in its interior. It could be due to the rotational dynamics of the superfluid in the neutron star's inner crust. The X-ray light curve can also get modulated due to release of large amount of X-ray energy during the sudden onset of glitches \citep{Hui04}. However, we have seen that the pulse profile is constant in time. There are no significant changes in the pulse shape parameters during glitches. Therefore, it is difficult to comment on the extent of turbulence caused by the glitches.

The underlying physics involved in the emission mechanism of Crab pulsar is not completely understood and coordinated multi-waveband study of the pulsar is crucial. Our results fill in the gaps between the trend seen at other wavelengths and we hope that observations by more sensitive instruments on future missions, like $ASTROSAT$ \citep{Paul09}, will provide important clues regarding the pulsar emission mechanism.

\normalem
\begin{acknowledgements}
We thank an anonymous referee for valuable comments that helped us to improve the paper.
We are grateful to Prof. W. Hermsen and Prof. L. Kuiper for helpful discussion on the analysis. This research has made use of data obtained from the High Energy Astrophysics Science Archive Research Center (HEASARC), provided by NASA's Goddard Space Flight Center. 
\end{acknowledgements}

\label{lastpage}
\end{document}